\documentclass[conference]{IEEEtran}
\usepackage[dvipdfmx]{graphicx}

\usepackage[caption=false,font=footnotesize]{subfig}

\usepackage{amsmath}

\usepackage{cite}

\usepackage{url}

\usepackage{siunitx}

\usepackage{lipsum}

\usepackage[mathscr]{euscript}

\usepackage{xcolor,colortbl}
\usepackage{color,soul}

\definecolor{morange}{rgb}{0.8,0.2,0}
\definecolor{mblue}{rgb}{0,0.3,1.0}
\definecolor{mpink}{rgb}{1.0,0.6,0.6}
\definecolor{mgreen}{rgb}{0.1,0.6,0.2}
\definecolor{mgoodgreen}{rgb}{0.9,1.0,0.7}

\definecolor{Gray}{gray}{0.85}

\usepackage{multirow}
\usepackage{array}
\newcolumntype{L}[1]{>{\raggedright\let\newline\\\arraybackslash\hspace{0pt}}m{#1}}
\newcolumntype{C}[1]{>{\centering\let\newline\\\arraybackslash\hspace{0pt}}m{#1}}
\newcolumntype{R}[1]{>{\raggedleft\let\newline\\\arraybackslash\hspace{0pt}}m{#1}}

\newcolumntype{G}{>{\columncolor{mgoodgreen}}c}

\begin{document}

 \title{Demo: A Unified Platform of Free-Space Optics for\\High-Quality Video Transmission}


\author{
\IEEEauthorblockN{Hong-Bae Jeon, Hyung-Joo Moon, Soo-Min Kim, Do-Hoon Kwon,\\ Joon-Woo Lee, Sang-Kook Han, and Chan-Byoung Chae}
\IEEEauthorblockA{Yonsei University, Korea\\ 
Email:\{hongbae08, moonhj, sm.kim, ehgns222, junu0809, skhan, cbchae\}@yonsei.ac.kr}
}


\maketitle

\begin{abstract}
In this paper, we investigate video signal transmission through an FPGA-based free-space optical (FSO) communication system prototype. We use a channel emulator that models the turbulence, scintillation, and power attenuation of the FSO channel and the FPGA-based real-time prototype for processing transmitted and received video signals. We vary the setup environment of the channel emulator by changing the amount of turbulence and wind speed. At the end of the demonstration, we show that through our testbed, even 4K ultra-high-definition (UHD) resolution video with 60 fps can be successfully transmitted under high turbulence and wind speed. \\

\end{abstract}

\IEEEpeerreviewmaketitle

\section{Introduction}
Recently, as the quality and size of videos have increased rapidly, a very high bit rate of the communication channel in video transmission, such as streaming services, has also been required~\cite{OCS}. The current radio frequency (RF) communication system, however, is now having difficulties solving the frequency licensing problem. As a result, the free-space optical (FSO) communication system with high capacity and license-free characteristics is emerging as a solution to construct a new high-bit-rate communication system~\cite{ACP}. 

One of the greatest weaknesses of FSO communication, however, is that the signal can easily fluctuate due to the channel characteristics, including atmospheric turbulence and scintillation. Hence, to combat the phenomena, robust real-time video transmission architectures with high bit rates have been studied in a variety of ways. In~\cite{VTOWC}, the author created a compact video transmission system with a channel-adaptive bit rate assuming the AWGN channel environment. In~\cite{JLT}, the author constructed a robust low-density-parity-check (LDPC) channel coding for the turbulent correlated FSO channels with intensity modulation and evaluated its performance by transmitting high-definition (HD) video signals. 

In this paper, we demonstrate real-time FPGA-based high-resolution video signal transmission via the FSO channel emulator, which models the channel characteristics of the FSO channel, including turbulence, scintillation, and power attenuation. We process the video signal via an electrical-to-optical (E/O) converter, an optical-to-electrical (O/E) converter, and the FPGA module and show that under harsh channel conditions, such as high turbulence or wind speed, our prototype successfully transmits and processes the video signal. More details and algorithms will be described in an extended draft. 

		\begin{figure}[t]
	\begin{center}
		\includegraphics[width=1\columnwidth,keepaspectratio]%
		{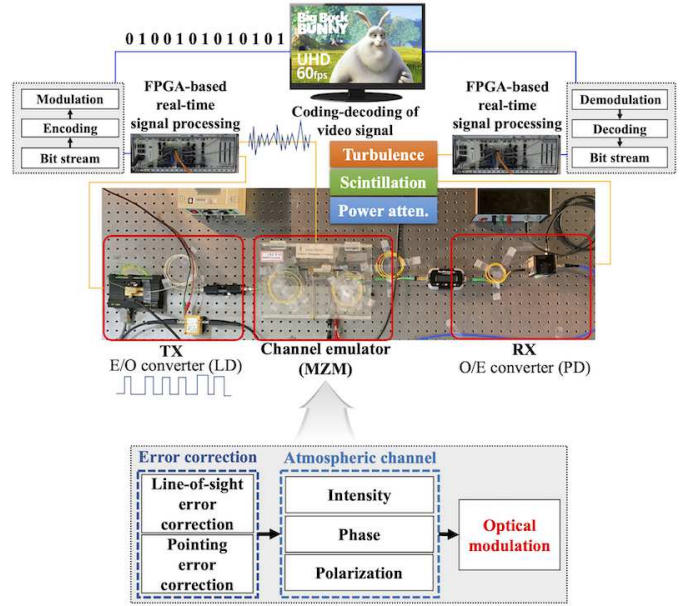}
		\caption{A prototype structure of video signal transmission according to parameter variation of atmospheric channel.}
		\label{Demo}
	\end{center}
\end{figure}
\begin{figure*}[t]
	\begin{center}
		\includegraphics[width=1.8\columnwidth,keepaspectratio]%
		{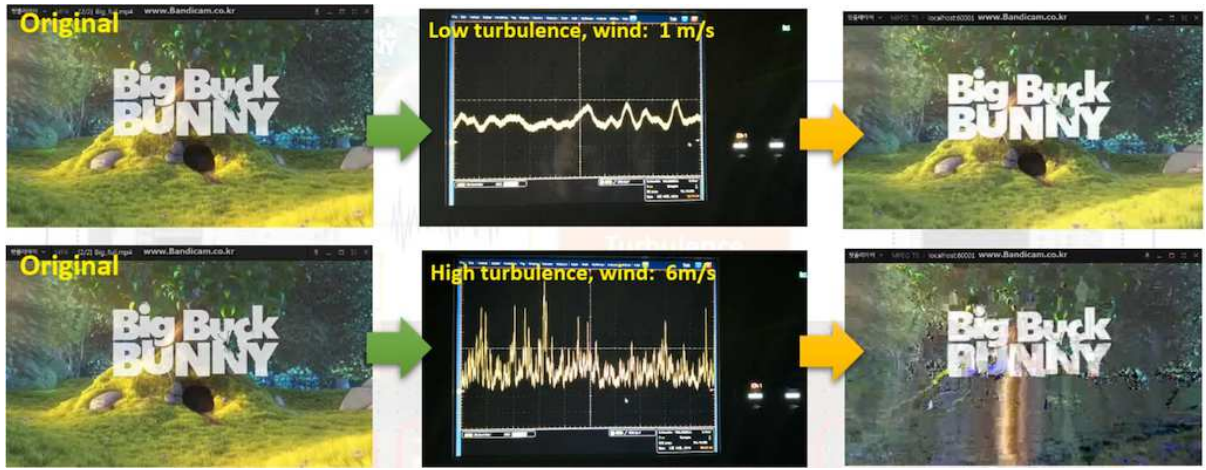}
		\caption{Snapshots of UHD video transmission results under several emulated FSO channel environments.}
		\label{Result}
	\end{center}
\end{figure*}
\section{System / Hardware Architecture}
\subsection{Hardware and Data Setup}
The prototype structure is shown in Fig.~\ref{Demo}. We used the Mach-Zehnder Modulator (MZM)~\cite{MZM} to model the atmospheric channel and the error correction, including both line-of-sight (LoS) and pointing errors. It is connected between the transmitter and receiver FPGA modules. The transmitter and receiver include the laser diode (LD) with an E/O converter and the photodiode (PD) with an O/E converter, respectively, which ensure long-term video signal transmission. The diodes are connected to the FPGA-based PXIe software-defined radio (SDR) platform for real-time video signal processing~\cite{SSC, NSIC}, which carries out the encoding and modulation of the bitstream of the video signal. The SDR platform at each transmitter and receiver side consists of a PXIe chassis (PXIe-1082) and FPGA controller modules (PXIe-8880, 8374, 2953R for transmitter and receiver side)~\cite{MK}. 

The type of transmitted data is a 1-minute ultra-high-definition (UHD: 3840 $\times$ 2160 resolution) 60 fps ``Big Buck Bunny” clip with H.264 codec. The bit rate is 200 kb/s and 48 kb/s for video and audio, respectively, and the sample rate is 22,050 Hz. For the FSO channel, we considered two conditions in the channel emulator: low turbulence with 1 m/s wind and high turbulence with 6 m/s wind. 

\subsection{Output Results}
Fig.~\ref{Result} shows the snapshots of the video transmission results under the channel conditions given in Section II.A. As we can see from the figure, under the environment with low turbulence and slow wind speed, the original clip is transmitted with almost zero distortion. Even under harsh conditions with high turbulence and fast wind speed, the clip is transmitted with low distortion. 


\section{Conclusion}
In this work, we developed a novel technique to model the real FSO channel with the MZM emulator and a point-to-point transmitter and receiver realized by FPGA modules. We assumed the FSO channel with both low and high turbulence scenarios. We showed that even a video signal with UHD 60 fps could be received without any or with less distortion by our SDR-platform-based prototype. We believe that more accurate video transmission techniques, which hold in practice, can be modeled and tested through our proposed testbed. For future work, we will focus on distortion removal techniques in the high turbulence scenario by using deep learning techniques.


\section*{Acknowledgment}
This work was partly supported by Institute for Information \& communications Technology Promotion (IITP) grant funded by the Korea government (MSIP) (No.2019-0-00685, Free space optical communication based vertical mobile network) and the Ministry of Science and ICT (MSIT), Korea, under the ``ICT Consilience Creative Program" (IITP-2019-2017-0-01015) supervised by the IITP.


\bibliographystyle{IEEEtran}
\bibliography{FSO}

\begin{thebibliography}{1}
\providecommand{\url}[1]{#1}
\csname url@samestyle\endcsname
\providecommand{\newblock}{\relax}
\providecommand{\bibinfo}[2]{#2}
\providecommand{\BIBentrySTDinterwordspacing}{\spaceskip=0pt\relax}
\providecommand{\BIBentryALTinterwordstretchfactor}{4}
\providecommand{\BIBentryALTinterwordspacing}{\spaceskip=\fontdimen2\font plus
\BIBentryALTinterwordstretchfactor\fontdimen3\font minus
  \fontdimen4\font\relax}
\providecommand{\BIBforeignlanguage}[2]{{%
\expandafter\ifx\csname l@#1\endcsname\relax
\typeout{** WARNING: IEEEtran.bst: No hyphenation pattern has been}%
\typeout{** loaded for the language `#1'. Using the pattern for}%
\typeout{** the default language instead.}%
\else
\language=\csname l@#1\endcsname
\fi
#2}}
\providecommand{\BIBdecl}{\relax}
\BIBdecl

\bibitem{OCS}
H.~{Kaushal} and G.~{Kaddoum}, ``Optical communication in space: Challenges and
  mitigation techniques,'' \emph{{IEEE} Commun. Surveys Tuts.}, vol.~19, no.~1,
  pp. 57--96, First Quarter 2017.

\bibitem{ACP}
M.~A. {Khalighi} and M.~{Uysal}, ``Survey on free space optical communication:
  A communication theory perspective,'' \emph{{IEEE} Commun. Surveys Tuts.},
  vol.~16, no.~4, pp. 2231--2258, Fourth Quarter 2014.

\bibitem{VTOWC}
S.~Ramesh, D.~Suresh, A.~Hussain, D.~Tamilvanan, and M.~P. Prabakaran, ``Video
  transmission through optical wireless communication,'' \emph{Int. J. Adv. R.
  Electr., Electron. Instrum. Eng. (IJAREEIE)}, vol.~2, no.~4, Apr. 2013.

\bibitem{JLT}
N.~{Cvijetic}, S.~G. {Wilson}, and R.~{Zarubica}, ``Performance evaluation of a
  novel converged architecture for digital-video transmission over optical
  wireless channels,'' \emph{J. Lightw. Technol.}, vol.~25, no.~11, pp.
  3366--3373, Nov. 2007.

\bibitem{MZM}
W.~M.~J. Green, M.~J. Rooks, L.~Sekaric, and Y.~A. Vlasov, ``Ultra-compact, low
  {RF} power, 10 {G}b/s silicon mach-zehnder modulator,'' \emph{Opt. Express},
  vol.~15, no.~25, pp. 17\,106--17\,113, Dec. 2007.

\bibitem{SSC}
J.~{Jang} \emph{et~al.}, ``Smart small cell with hybrid beamforming for 5g:
  Theoretical feasibility and prototype results,'' \emph{IEEE Wireless
  Commun.}, vol.~23, no.~6, pp. 124--131, Dec. 2016.

\bibitem{NSIC}
M.-S. {Sim}, M.~{Chung}, D.~{Kim}, J.~{Chung}, D.-K. {Kim}, and C.-B. {Chae},
  ``Nonlinear self-interference cancellation for full-duplex radios: From
  link-level and system-level performance perspectives,'' \emph{IEEE Commun.
  Mag.}, vol.~55, no.~9, pp. 158--167, Sep. 2017.

\bibitem{MK}
M.~{Chung}, M.-S. {Sim}, J.~{Kim}, D.-K. {Kim}, and C.-B. {Chae}, ``Prototyping
  real-time full duplex radios,'' \emph{IEEE Commun. Mag.}, vol.~53, no.~9, pp.
  56--63, Sep. 2015.

\end{thebibliography}
\end{document}